# Fuzzy Logic Based Integration of Web Contextual Linguistic Structures for Enriching Conceptual Visual Representations

M. Belkhatir, University of Lyon I

*Abstract*—Due to the difficulty of automatically mapping visual features with semantic descriptors, state-of-the-art frameworks have exhibited poor performance in terms of coverage and effectiveness for indexing the visual content. This prompted us to investigate the use of both the Web as a large information source from where to extract relevant contextual linguistic information and bimodal visual-textual indexing as a technique to enrich the vocabulary of index concepts. Our proposal is based on the Signal/Semantic approach for multimedia indexing which generates multi-facetted conceptual representations of the visual content. We propose to enrich these image representations with concepts automatically extracted from the visual contextual information. We specifically target the integration of semantic concepts which are more specific than the initial index concepts since they represent the visual content with greater accuracy and precision. Also, we aim to correct the faulty indexes resulting from the automatic semantic tagging. Experimentally, the details of the prototyping are given and the presented technique is tested in a Web-scale evaluation on 30 queries representing elaborate image scenes.

## I. INTRODUCTION

The World Wide Web, in less than two decades, has transformed the way people communicate and socialize. Digital cameras and cheap storage create an abundance of digital information, not to mention the rise of the Web 2.0 that is enabling people to upload and share visual documents and other multimedia contents on the Web.

Contextual information, an important feature of Web images, is especially attractive to automatically characterize the image content as the success gained by Web image search engines has shown its relative effectiveness. Multimodal systems which fuse both visual content and contextual textual features of Web images, has demonstrated the need for improved textual characterization and processing as the inaccuracy of the textual information attached to Web images degrades the effectiveness of their automated characterization [1]. As an improvement, data gathered from Web 2.0 websites have been used to enhance the performance of concept-based retrieval systems. In particular, these websites put a specific emphasis on tags for automatically characterizing and searching the multimedia content.

However, Yang et al. [2] pinpoint the insufficiency of tags in providing a complete description of the document content. They therefore propose to enrich tag descriptions with additional visual properties. In the effort to enrich tag characterization, Ruocco and Ramampiaro [3] propose a method based on statistical exploratory analysis of spatial point patterns to extract spatial features from image tags. These are used both to summarize spatial distributions of single tags and to determine spatial relatedness between couples of tags. The issue of noisy tags (i.e. tags that inaccurately characterize the multimedia content) has been addressed by Tang et al. [4] and Wang et al. [5] to highlight the semantic content of Web 2.0 and facial images respectively. Larson et al. [6] propose themselves to enrich the informativeness of image tags, i.e. how related they are to the image content, by combing a measure predicting whether the tags correspond to a physical entity and its associated size (i.e. small, medium or large) as well as statistics of web-mined natural language. Tian et al. [7] incorporate semantics and linguistics techniques to filter out irrelevant tags, retaining only object and scene tags with pair-wise semantic similarity. The tag set is then enriched with verbs and content-based visual contextual tags. Sun et al. [8] consider themselves humans and objects and their interrelationships within a Bayesian framework for image retagging. Tags illustrating the mutual context of human and objects are linked to those interrelationships through a probabilistic graphical model.

The main issue of state-of-the-art proposals relies on the fact that the information generated from the different modalities is gathered in a typical bag-of-words approach, which is unsufficient for the processing of more linguistically complex queries, i.e. queries that do not consist only of trivial tags, as is sometimes experienced when using the Web based image search engines. But if used with both the visual content and semantic concepts derived from the visual content within a theoretically sound integrated framework, the relevance and precision of automated image characterization and search effectiveness can be further improved.

In order to enrich the semantic characterization of the state-of-the-art image description frameworks, we develop a theoretically sound conceptual model taking into account the contextual visual information in webpages as an enrichment source. Concepts and relations describing the visual content are automatically extracted and then integrated in its description. The main goal is to enrich visual image representations with concepts and relations which are more precise and specific than the ones obtained through the first process of semantic characterization. Also, since the latter process is error-prone, an other major aim will consist of correcting the potential erroneous semantic descriptors.

The visual indexing framework revolves around the concept of visual object (VO), abstraction representing a visual entity in a document. At the core of our proposal is the notion of **Visual Index Structure** (VIS), a text-based structure describing the VOs in an image, their semantic, color and texture descriptions (noted respectively $s\_d$, $c\_d$ and $t\_d$ in the grammar of fig. 1) as well as the spatial descriptions (noted $spa\_d$ in the grammar of fig. 1) linking them.

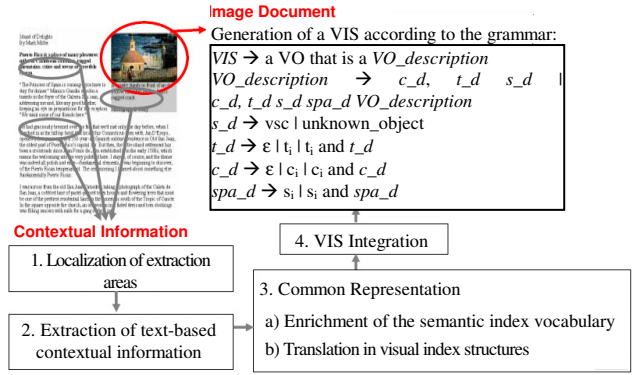

Fig. 1. Overview of the framework

Accordingly, we summarize the different steps of the research. We first highlight and extract the information relevant to the Web image document (step 1 of fig. 1). Syntactic patterns corresponding to the index structures that describe the visual content are defined to extract the contextual information (step 2 of fig. 1). We then define a common representation frame for elements extracted from the Web on the one hand and the visual index structures on the other hand (step 3 of fig. 1), allowing the enrichment of the semantic index description (step 4 of fig. 1). We particularly investigate tools from fuzzy set theory and semantic technology. Experimentally, we propose a system-based evaluation on a dataset of Web images and their context.

## II. CONCEPTUAL VISUAL INDEXING

A VO is assigned a visual semantic concept $vsc$ which is learned and then automatically extracted given a visual semantic vocabulary $V_{sem}$ through semantic concept detectors (cf. [9] for the details). A posterior recognition probability value $r_{vsc}$ is computed which measures the certainty that $vsc$ describes the indexed VO. Visual semantic concepts are organized within a concept lattice $Lat_{vsc}$ ordered by a hypernymy/hyponymy relation (shown in fig. 2).

The signal color description of a VO is given through eleven automatically-assigned color concepts with their corresponding pixel percentage (cf. [10] for the description). The color vocabulary $V_c$ consists of color concepts: $c_1$=cyan(C), $c_2$=white(W), $c_3$=green(Gn), $c_4$=grey(G), $c_5$=yellow(Y), $c_6$=black(B), $c_7$=orange(O), $c_8$=skin(S), $c_9$=red(R), $c_{10}$=blue(Bl) and $c_{11}$=purple(P).

The signal texture description of a VO is characterized through a texture vocabulary $V_{tx}$ consisting of eleven texture concepts: $t_1$=bumpy(B), $t_2$=cracked(C), $t_3$=disordered(D), $t_4$=interlaced(I), $t_5$=lined(L), $t_6$=marbled(M), $t_7$=netlike(N), $t_8$=smeared(S), $t_9$=spotted(Sp), $t_{10}$=uniform(U) and $t_{11}$=whirly (W). They are automatically extracted with an SVM-based framework (cf. [11] for its characterization).

We propose a spatial vocabulary $V_{spa}$ consisting of eleven concepts that indicate the relative positions of VOs within a document. Considering 2 VOs (Vo1 and Vo2), these relations are ($s_1$=C,Vo1,Vo2): 'Vo1 partially covers (in front of) Vo2', ($s_2$=C_B,Vo1,Vo2): 'Vo1 is covered by (behind) Vo2', ($s_3$=P,Vo1,Vo2): 'Vo1 is a part of Vo2', ($s_4$=T,Vo1,Vo2): 'Vo1 touches Vo2 (is externally connected)' and ($s_5$=D,Vo1,Vo2): 'Vo1 is disconnected from Vo2'. Directional relations Right($s_6$=R), Left($s_7$=L), Above($s_8$=A), Below($s_9$=B) are invariant to basic geometrical transformations (translation, scaling). Two relations specified in the metric space are based on the distances between VOs. They are the Near($s_{10}$=N) and Far($s_{11}$=F) relations. All are automatically computed as explained in [12].

Once the color, texture and spatial concepts are derived, a VIS is automatically generated according to the context-free grammar in fig. 1.

## III. CONTEXTUAL CONTENT MODELING

The first task consists in defining the area from which the contextual linguistic information in the webpage is extracted (step 1 of fig. 1). We use the alt and src attributes of the image link as well as the surrounding text. For each text extraction area in the webpage, we define a value $imp(cx) \in [0,1]$ characterizing the probability that a contextual concept $cx$ extracted from this area is relevant to the visual content. When a concept occurs several times, its value $imp(cx)$ is the maximum of all impact values computed for each of its occurrences.

Banking on the hypothesis that any image document is represented by a VIS, the goal of our work is to relate the Web linguistic components to this representation. For this, we extract concepts from the specified locations then define the syntactic terms that characterize the visual entities while being at the same time related to the visual semantic, color, texture and spatial concepts of the VIS. For this, we define specific syntactic patterns (step 2 of fig. 1).

A syntactic pattern is defined as a list of concepts extracted from the contextual information that belong to one of these syntactic categories: P=[$cx_0$, $cx_1$, ... $cx_n$] where: $cx_i \in V_{sem\_enr} \cup V_c \cup V_{tx} \cup V_{spa} \cup other$. The vocabulary $V_{sem\_enr}$ consists of the semantic concepts in $V_{sem}$ enriched with concepts extracted from the contextual information (they are semantically related to concepts in $V_{sem}$ as explained in section IV.A). The set *other* consists of all other concepts which can potentially occur in between the concepts of the vocabularies considered. For example, the extracted string "vegetation scene with red flowers" is represented by the pattern [$cx_0$, $cx_1$, $cx_2$, $cx_3$, $cx_4$] where $cx_0 \in V_{sem\_enr}$, $cx_1 \in$ other, $cx_2 \in$ other, $cx_3 \in V_c$, $cx_4 \in V_{sem\_enr}$ (indeed concepts "vegetation" and "flower" are comprised in $V_{sem\_enr}$ as shown in fig. 2 and the color concept "red" belongs to $V_c$ while concepts "scene" and "with" do not belong to either $V_{sem\_enr}$, $V_c$, $V_{tx}$ or $V_{spa}$).

Applying patterns to the text extracted from webpages results in a set of syntactic terms conform to a VIS. Each syntactic term *st* is a set of pairs such that st={(cx, imp(cx))} with *cx* a concept extracted from the contextual information and *imp(cx)* its impact value given its position in the webpage (i.e. whether it is found in the alt, src attributes or in the surrounding text). Indeed, since we consider the relevance of a concept with respect to the visual content, its impact value expresses its probability of relevance given its location of extraction. Let us note that the cardinality of *st* depends on the applied pattern.

## IV. ENRICHING VISUAL INDEX STRUCTURES

To establish a correspondence between a syntactic term and the VIS of a VO, we first define a common representation frame (step 3 of fig. 1). For this, we propose an enriched lattice of visual semantic concepts which integrates the new semantic concepts extracted from the contextual information in webpages. We then define functions to estimate structure similarity and finally propose a fusion scheme (step 4 of fig.1) to refine the accuracy and precision of the semantic characterization of VISs.

### A. Enriched Lattice of Semantic Concepts

To build an enriched lattice of concepts corresponding to the elements of the syntactic terms, we use the visual semantic concept lattice $Lat_{vsc}$ in which we add the concepts extracted from the contextual information. The integration of these additional concepts in the lattice is achieved through the specific/generic partial order $\leq_{sem}$ (translated as the *is_a* relation, for example: a cathedral is a building). This process results in an enriched lattice $Lat_{vsc\_enr}$ of semantic concepts (a part of it is represented in fig. 2).

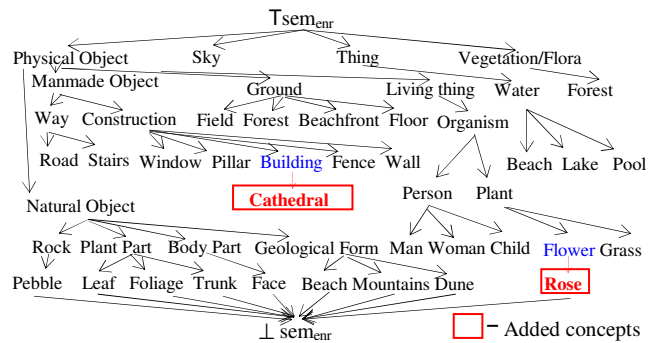

Fig. 2. A Part of the Enriched Lattice of Semantic Concepts

In fig. 2, we show the integration of two new semantic concepts: "rose" and "cathedral" respectively as concepts specific to concepts "flower" and "building" initially in $Lat_{vsc}$.

### B. Structure Similarity

We determine the similarity between a VIS and a syntactic term through the use of fuzzy sets, which are a generalization of set theory with a membership relation transformed in a function with values in the interval [0,1], by taking into account three factors: i) the similarity (with respect to the enriched lattice of visual semantic concepts $Lat_{vsc\_enr}$) between their semantic concepts; ii) the impact of the concepts derived from the syntactic terms with respect to their position in the webpage; iii) the recognition probability value of a visual semantic concept in the VIS. Let $U_{sem}$ the (discrete) universe of all semantic concepts. We define 2 membership functions characterizing the likelihood that a concept $c \in U_{sem}$ describes the visual content.

The membership function $\mu_{cx}(c):U_{sem} \rightarrow [0,1]$ associates to an element *c* of the universe $U_{sem}$ the likelihood that it describes a visual entity in the document also described by the contextual concept cx:

$$\mu_{cx}(c) = \begin{cases} imp(cx) & \text{if } c \text{ is a generic or equal to cx with respect to } Lat_{vsc\_enr} \\ imp(cx)+path\_length(cx,c) & \text{if } c \text{ is a specific of cx} \\ 0 & \text{otherwise} \end{cases} \quad (1)$$

It highlights the fact that the impact value *imp(cx)* is propagated to the most generic concepts. If the concept *c* is more specific than *cx*, the goal is then to reinforce its importance. The degree of specificity between concepts is quantitatively characterized by the *path_length* function in the enriched lattice of semantic concepts normalized by the longest path [10]. This value is then added to the impact value of cx (i.e. *imp(cx)*) to derive $\mu_{cx}(c)$. In the case where there is no semantic correspondence between concepts *cx* and *c* (i.e. in terms of the *is_a* relation), we assign to *c* a null membership value.

We also define the membership function $\mu_{vsc}(c):U_{sem} \rightarrow [0,1]$, which associates to a concept *c* of the universe $U_{sem}$ the likelihood that it describes the visual content indexed by the semantic concept *vsc* in the VIS:

$$\mu_{vsc}(c) = \begin{cases} r_{vsc} & \text{if } c \text{ is a generic or equal to vsc with respect to } Lat_{vsc\_enr} \\ r_{vsc}+path\_length(c,vsc) & \text{if } c \text{ is a specific of vsc} \\ 0 & \text{otherwise} \end{cases} \quad (2)$$

Following the same modus operandi as in the previous case, we propagate the probability recognition value $r_{vsc}$ to the most generic concepts. Moreover, we also favor the specific concepts through the *path_length* function.

## C. Reconciliation

We estimate the effect of all concepts from the VIS by the membership function $\mu_{tot\_vis}(c)$ which is the result of aggregating values $\mu_{vsc}(c)$ of a concept $c$ for all visual semantic concepts of $V_{sem}$. Similarly, the function $\mu_{tot\_cx}(c)$ is computed as the result of the aggregation of values $\mu_{cx}(c)$ for all concepts extracted from the contextual information. The aggregation of the membership functions is achieved through a t-conorm: commutative and associative application with increasing values from $[0,1]^2$ to $[0,1]$. Accordingly, we consider the contribution of all concepts semantically related to $c$, which reinforces the fact that the latter describes the visual content. The algorithm used to derive these values is detailed in section V.A.

Values $\mu_{tot\_vis}(c)$ and $\mu_{tot\_cx}(c)$ are then fused to compute the likelihood that the concept $c$ describes the visual content by taking into account all other concepts. For all $c \in U_{sem}$, we aggregate the two values $\mu_{tot\_vis}(c)$ and $\mu_{tot\_cx}(c)$ to obtain a membership value $\mu_{tot}(c)$ which evaluates the likelihood that the concept $c$ characterizes the visual content by considering the impact of the visual semantic concepts in the VIS on the one hand and the contextual concepts on the other hand.

The algorithm for generating the $\mu_{tot}$ values is given below:

```
procedure compμ_tot
  var : list_of_concepts in U_sem, V_sem, V_sem_enr
  begin
    for i=0 to length(U_sem)
      μ_tot_vis[i] = comp_μ_t(U_sem[i], V_sem)
      μ_tot_cx[i]  = comp_μ_t(U_sem[i], V_sem_enr)
      μ_tot[i]     = S2(μ_tot_vis[i], μ_tot_cx[i])
  end

function comp_μ_t (var1: (concept,value) pair, var2: list of (concept, value) pairs)
/*for computing formulas (1) and (2) with WordNet (cf. IV.B)*/
  var : μ_tot_v(0)=0
  begin
    for j=1 to length(var2)
      if var1.concept hypernym of var2[j].concept then
        μ_tot_v(j)=var2[j].value
      else if var1.concept hyponym of var2[j].concept then
        μ_tot_v(j)=min((var2[j].var+path(var1.concept,var2[j].concept)),1)
        /* path is the similarity function provided in WordNet */
      else if not (var1.concept synonym of var2[j].concept) then
        μ_tot_v(j)=0
      μ_tot_v(j)=S2(μ_tot_v(j), μ_tot_v(j-1))
    return μ_tot_v
  end
```

### 1) Estimating the similarity

We define a value estimating the similarity between a VIS $vis$ and a syntactic term $st$ given by formula (3):

$$\mathrm{sim}(st, vis) = \sum_{j=1}^{Card(V_{tx})} \frac{\max(T_{st}[j], T_{vis}[j])}{Card(V_{tx})} + \sum_{j=1}^{Card(V_{spa})} \frac{\max(S_{st}[j], S_{vis}[j])}{Card(V_{spa})} + \sum_{j=1}^{Card(V_c)} \frac{\max(C_{st}[j], C_{vis}[j])}{Card(V_c)} + \quad (3)$$
$$\varepsilon(vsc, cx) \cdot [\mu_{tot}(vsc) + \mu_{tot}(cx)]$$

Here, $T_{st}$, $S_{st}$ and $C_{st}$ are vector structures translating respectively the signal texture, spatial and color characterizations in $st$. Values $T_{st}[j]$, $S_{st}[j]$ and $C_{st}[j]$ respectively quantify texture concept $t_j$, spatial concept $s_j$ and color concept $c$. In the same fashion, $T_{vis}$, $S_{vis}$ and $C_{vis}$ translate the signal texture, spatial and color characterizations in $vis$.

By considering summing their correspond values at index $j$, we reinforce the similarity between two structures sharing the same attributes as far as their signal color, texture and spatial characterizations are concerned. We normalize this sum by the cardinality of the texture, spatial and color vocabularies so as not to introduce a strong bias in the case where their signal descriptions do not correspond. The semantic correspondence between $st$ and $vis$ is computed by considering the function $\varepsilon$ which estimates the similarity between the two concepts $vsc$ and $cx$ with respect to the lattice $Lat_{vsc\_enr}$, weighted by their fuzzy values $\mu_{tot}(vsc)$ and $\mu_{tot}(cx)$.

### 2) Study of similarity cases

We compute all similarity values $sim_{i,k}$ between syntactic terms $st_i$ and VISs $vis_k$. A syntactic term does not necessarily correspond to a unique VIS and vice versa. We distinguish 3 cases of correspondence:

▪ Syntactic term $st_i$ corresponds to a unique VIS $vis_k$, i.e. we have highlighted a similarity between the semantic concepts of the two structures.
▪ Syntactic term $st_i$ corresponds to $n_{vis}$ VISs. In this case, the semantic concept of the syntactic term corresponds to the semantic concepts of the $n_{vis}$ VISs.
▪ $n_{st}$ syntactic terms correspond to a unique VIS $vis_k$, i.e. the $n_{st}$ semantic concepts in the syntactic terms correspond to a unique semantic concept in $vis_k$.

We highlight these different cases of correspondence and the similarity values $sim_{ik}$ between structure pairs in fig. 3 which illustrates a bipartite graph $G_{st\_vis}$ with $n_{st}$ and $n_{vis}$ denoting the number of nodes. The objective is to determine the pairs of structures achieving the best correspondence, i.e. the pairs $(st_i, vis_k)$ for which the similarity values $sim_{i,k}$ are maximum so as to integrate them in the visual index description. The result of this step is a set of structure pairs achieving the best correspondence.

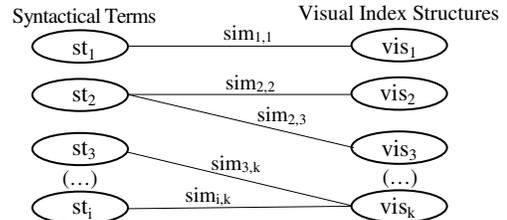

Fig. 3. Cases of correspondence between syntactical terms $st_i$ and the visual index structures $vis_k$

### 3) Structure Fusion

In this last step, we integrate the elements extracted from the Web contextual content in the visual description. We shall now consider fusing these structures according to a two-step process:

▪ Decide whether to replace *vsc*, the visual semantic concept of the VIS, with *cx*, the contextual concept of the syntactic term. This is done only in the case where *cx* is more specific than *vsc* or the latter is faulty and needs to be corrected.
▪ Reinforce the fuzzy value of the chosen semantic concept which translates to what extent the concept describes the visual content.

To determine the choice of the concept to be kept in the VIS, we compare the $\mu_{tot}$ values of the two concepts. We define a threshold $T_\mu$ to quantify the difference between the two membership values and highlight a semantic correspondence between the two concepts. In the case of a semantic correspondence, i.e. $|\mu_{tot}(vsc)-\mu_{tot}(cx)| \leq T_\mu$, we consider the relationship between *vsc* and *cx*. If *vsc* is more specific than *cx*, the first is not modified. If *cx* is more specific than *vsc*, then the latter is replaced with the first. If there is no semantic correspondence, i.e. $|\mu_{tot}(vsc)-\mu_{tot}(cx)|>T_\mu$, if $\mu_{tot}(vsc)-\mu_{tot}(cx)<0$, we keep the concept *vsc* in the VIS. Otherwise, we replace it with concept *cx*.

Afterwards, we compute the likelihood that the chosen concept describes the visual content. It is equal to the maximum value between $\mu_{tot}(vsc)$ and $\mu_{tot}(cx)$.

## V. EXPERIMENTAL INSTANTIATION

### A. Implementation of the framework

The enrichment process starts with the automatic acquisition of WWW contextual contents containing at least an image with dimensions respecting predefined constraints [13] [14]. The automatic content-based indexing is handled by the framework which highlights visual objects, automatically assign their semantic concepts and generate a visual index structure, for a given document. The consecutive steps of the framework implementation for enriching the visual index structures as defined in the theoretical section are illustrated in fig. 4.

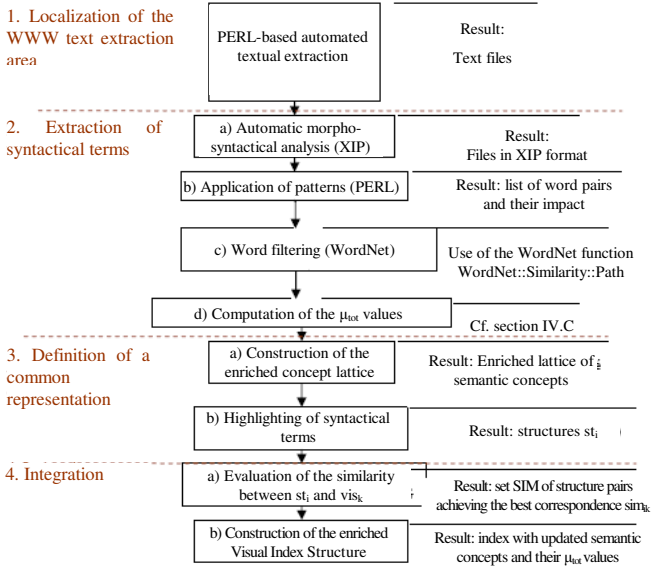

Fig. 4. Implementation steps of the proposed framework

The extraction of the contextual text-based information from the locations previously specified is automatic and handled by the XIP software to obtain its morpho-syntactical analysis. Patterns for extracting syntactical terms (introduced in section III) are implemented as regular expressions in the experimental setting. Elements of the vocabulary are then filtered to eliminate terms which do not correspond to semantic concepts and their relations. We use WordNet as the ontology to estimate the similarity between elements of the visual indexing vocabulary and those of the Web visual contextual information. The chosen similarity measure is the WordNet path which provides values in the [0,1] interval. For color, texture adjectives and spatial verbs and adverbs extracted from the contextual information, we check whether they respectively correspond to color, texture categories and spatial relations of the VIS.
WordNet ontology is used to generate the enriched lattice of semantic concepts which are related through hyponymy/hypernymy relations.

### B. Web-Scale Evaluation on Topic-based Queries

For the evaluation of the semantic enrichment resulting from the integration of the Web contextual information within the visual index structures, we propose a web-scale evaluation where one popular image search engine is adopted as the baseline search engine. We use it to retrieve images representing elaborate image scenes through topic-based queries which involve multiple characterizations of the visual content, as opposed to tag-based queries. We therefore specify thirty queries involving semantic concepts with color, texture and spatial characterizations in table 1.

TABLE I
Queries involving multiple facets of the visual content

| Smeared Walls | Green and White Walls | Lined Flowers |
|---|---|---|
| Lined People | Spotted Floors | People in front of buildings |
| Water near sand | Black and White Sky | Mountain view from far |
| People near buildings | Smeared Sand | Brick Ground |
| Sandy Ground | Vegetation above Ground | Netlike Window |
| Bumpy Walls | White and Red Tower | Smeared Buildings |
| People outside Water | Disordered/Covered Sky | Smeared Roads |
| People inside pool | Interlaced Foliage | Smooth Sky |
| Whirly Water | Bumpy Road | People near foliage |
| Uniform Floor | Whirly Flowers | Uniform Crowd |

Up to 500 returned images for each query are crawled to construct the experimental data set. To obtain the ground truth of the relevance orders of the returned images, we resorted to a manual labeling procedure. We evaluate the performance of six indexing strategies:

• II: baseline technique where images are searched using the initial index (used by the search engine);
• Sem: searching images using a state-of-the-art semantic-based visual indexing technique [14];
• Tf-Idf: words of the processed contextual information are weighted according to a *tf-idf* scheme. The latter has been used in text retrieval to estimate the importance or relevance of a word based on frequency attributes [14];
• VIS: searching images using the original VIS [15];

- CX: searching images using the syntactic terms extracted from the contextual information as highlighted in section 3;
- VIS+CX: searching images using the VIS enriched with the contextual information as described in section 4;

We adopt the Normalized Discounted Cumulative Gain at n (NDCG@n) as the evaluation metric which is especially suited in the case of Web search evaluation. Contrarily to other standard metrics that take into account experimental settings with binary relevance judgments, NDCG@n considers several degrees of relevance according to:

$$NDCG@n = Norm_n \sum_{i=1}^{n} \frac{2^{rlv_i} - 1}{\log(i+1)} \quad (4)$$

where $Norm_n$ is is a normalization constant such that the 'perfect' retrieval list has an NDCG value equal to 1 and $rlv_i$ is the relevance score of the $i^{th}$ image in the retrieval list. This metric favors retrieval lists where image results are sorted according to query relevance. This is consistent with users' expectation that images relevant to their queries are retrieved at the top positions of the retrieval results. Fig. 5 shows the performance of the compared approaches.

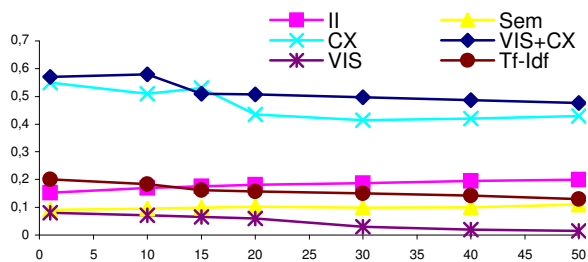

Fig. 5. Comparison of NDCG@k for search with 6 indexing schemes

The *II* and *Sem* techniques lead to unsatisfying performance due respectively to the ambiguity of the initial query for the first and not taking into account the contextual text information for the second. The poor performance of the *Tf-Idf* indexing strategy can be explained by the fact that words are equally weighted regardless of their position. Indeed, the position of a word with respect to the image has an effect on its relevance with respect to the visual content. Also, contextual information that precisely describes the image might consist of words with low frequency yet still relevant to the image content. *CX* which considers indexing with contextual information performs much better thanks to the localization of the information relevant to the visual content. Finally, by further refining the VIS using contextual information, the VIS+CX strategy obtains the best performance.

## VI. CONCLUSION

We have used the visual contextual information of multimedia documents gathered from the Web in order to enrich the semantic characterization of their visual content. Our work is based on the extraction of syntactic terms, elements which characterize semantic concepts with their visual attributes and their transformation into conceptual structures. The latter are then integrated in the visual index representation which describes the visual content through a multi-facetted conceptual index structure. The result of this integration is the representation of the semantic content through concepts which are more specific than the initial visual semantic concepts. Moreover, the new visual description is reinforced by the fact that a given semantic concept is highlighted in the Web visual context, which could resolve ambiguities introduced by the recognition and automatic tagging of semantic concepts. This last aspect of our aims at correcting erroneous indexes brought about by visual processing-based automatic indexing processes. Further developments will consist in enriching the knowledge resources for indexing [16].